\documentclass[twocolumn,prl,amsmath,amssymb,showpacs,superscriptaddress,floatfix]{revtex4}
\usepackage{graphicx}% Include figure files
\usepackage{bm}% bold math
\usepackage{amsmath}

\usepackage{epstopdf}
\usepackage{epsf}
\begin{document}
%\draft
\title{Interplay between freezing and density anomaly in a confined core-softened system}

\author{Yu. D. Fomin}
\affiliation{Institute for High Pressure Physics, Russian Academy
of Sciences, Troitsk 108840, Moscow, Russia}
\author{V. N. Ryzhov}
\affiliation{Institute for High Pressure Physics, Russian Academy
of Sciences, Troitsk 108840, Moscow, Russia}
\author{E. N. Tsiok}
\affiliation{Institute for High Pressure Physics, Russian Academy
of Sciences, Troitsk 108840, Moscow, Russia}

\begin{abstract}
We present a computer simulation study of influence of the confinement on the density anomaly in the system with isotropic core-softened potential which is used for a qualitative description of the anomalous behavior of water and some other liquids. We have found that a maximum temperature of the density anomaly region  along an isochor $\rho=0.5$ does not depend on the pore width. We have shown that a decrease in the width of the confining slit pore leads to an increase in the crystallization temperature and as a result the local density distribution is mostly controlled  by the particle-wall interaction, which leads to the depression of the density anomaly.
\end{abstract}


\maketitle

It is well known that some liquids can demonstrate anomalous behavior, i.e. the behavior which is not typical for
most of substances. The most well known example of an anomalous liquid is water which demonstrates more then seventy anomalies \cite{wateranom}.
The most vividly discussed anomaly of water is the density anomaly, i.e. ability of water to shrink upon heating or, equivalently,
negative value of thermal expansion coefficient $\alpha_P$. From the relation $\left( \frac{\partial P}{\partial T} \right)_V=\alpha_P/K_T$, where
$K_T$ is the isothermal compressibility, one can see that density anomaly corresponds to decreasing of pressure with temperature along isochors,
and the point of minimum gives the boundary of the region of density anomaly.

In spite of numerous works, the origin of anomalous properties of water remains vague \cite{tale}. Several models were proposed
to explain complex behavior of water. Among them the model which propose that the second critical point exists in strongly
supercooled region is the most popular at the moment \cite{cp2}. This model predicts that a liquid-liquid phase transformation takes place
in strongly supercooled water. Unfortunately, this assumption is difficult to verify experimentally, since
the region of this phase transition is separated from ambient conditions by so-called 'no man's land' - a region of pressure and
temperature values which cannot be reached in experiment due to crystallization.

A possible method to stabilize liquid water to cross the no man's land and reach the liquid-liquid transition line is to use water
confined in some porous medium (see, for instance, \cite{mallomaco,conf-1} and references therein). However,
the behavior of confined liquids can be strongly different from the bulk ones. Because
of this it becomes of particular importance to monitor the influence of the confinement on the anomalous properties of water and
in particular the influence of confinement on the density anomaly of water.

Although water is the most widespread liquid, it appears to be very complex system to study. In particular, until now
there is no any computer model which can reproduce many different properties of water simultaneously \cite{vega-comparison,vega-comparison1}.
Because of this it makes sense to study not only water, but also simpler systems which also demonstrate anomalous properties.

Many anomalies can be found in so-called 'core-softened systems', i.e. the systems in which the repulsive core of interaction potential is
softened. Many core-softened systems demonstrate some water-like anomalies (density anomaly, diffusion anomaly, structural anomaly, etc.)
\cite{cores1,cores2,cores3,cores4,cores5,cores6,cores7,cores8}. Because of this investigation of core-softened systems
gives an opportunity to study the qualitative behavior of anomalous liquids in systems which are computationally cheaper than real water.

In the present work we investigate the density anomaly in a core-softened system confined in a slit pore. We study a core-softened system with a potential:

\begin{equation}\label{pot}
  U(r)/ \varepsilon= \left( \frac{\sigma}{r} \right)^{14}+0.5 \left(
  1-tanh(k\cdot(r-\sigma_1) \right),
\end{equation}
where $\varepsilon$ and $\sigma$ give the scales of energy and
length, respectively. Below all quantities will be expressed in the units based
on these scales, for example the density $\tilde{\rho}=N/V*
\sigma^3$. Since only these reduced units are used we omit the
tilde marks. The parameters are $k=10.0$, $\sigma_1=1.35$. The
cut-off distance is $r_c=2.5$ This system was studied in details
in many recent publications \cite{we1,we2,we3,we4,we5,we6,we7,we8,we9,we10,we11}. In particular, it was shown
that the phase diagram of the system is very complex. It
includes numerous crystalline phases. The crystalline phase with
the lowest density is a Face centered cubic (FCC) one. It
demonstrates a maximum on the melting line. At higher densities
there is a region where no stable crystalline phase was
discovered. However, it was shown that the system in this region
characterized by extremely slow dynamics and demonstrated a
glass transition \cite{we1,ryltsev-glass}. Also at the densities slightly above the low
density FCC a set of dynamic and
thermodynamic anomalies was found: diffusion anomaly (the diffusion
coefficient increases under isothermal compression), density
anomaly (negative values of the thermal expansion coefficient),
structural anomaly (increasing of the excess entropy under
isothermal compression), maxima at the dependencies of the
response functions (isobaric heat capacity, thermal
compressibility) \cite{we2,we3,we4,we5,we6,we7,we8,we9,we10,we11}. Also, the system shows anomalously high
magnitude of the isochoric heat capacity due to smooth crossover
of the fluid structure \cite{cv1,cv2}.

\begin{figure}
\includegraphics[width=6cm, height=6cm]{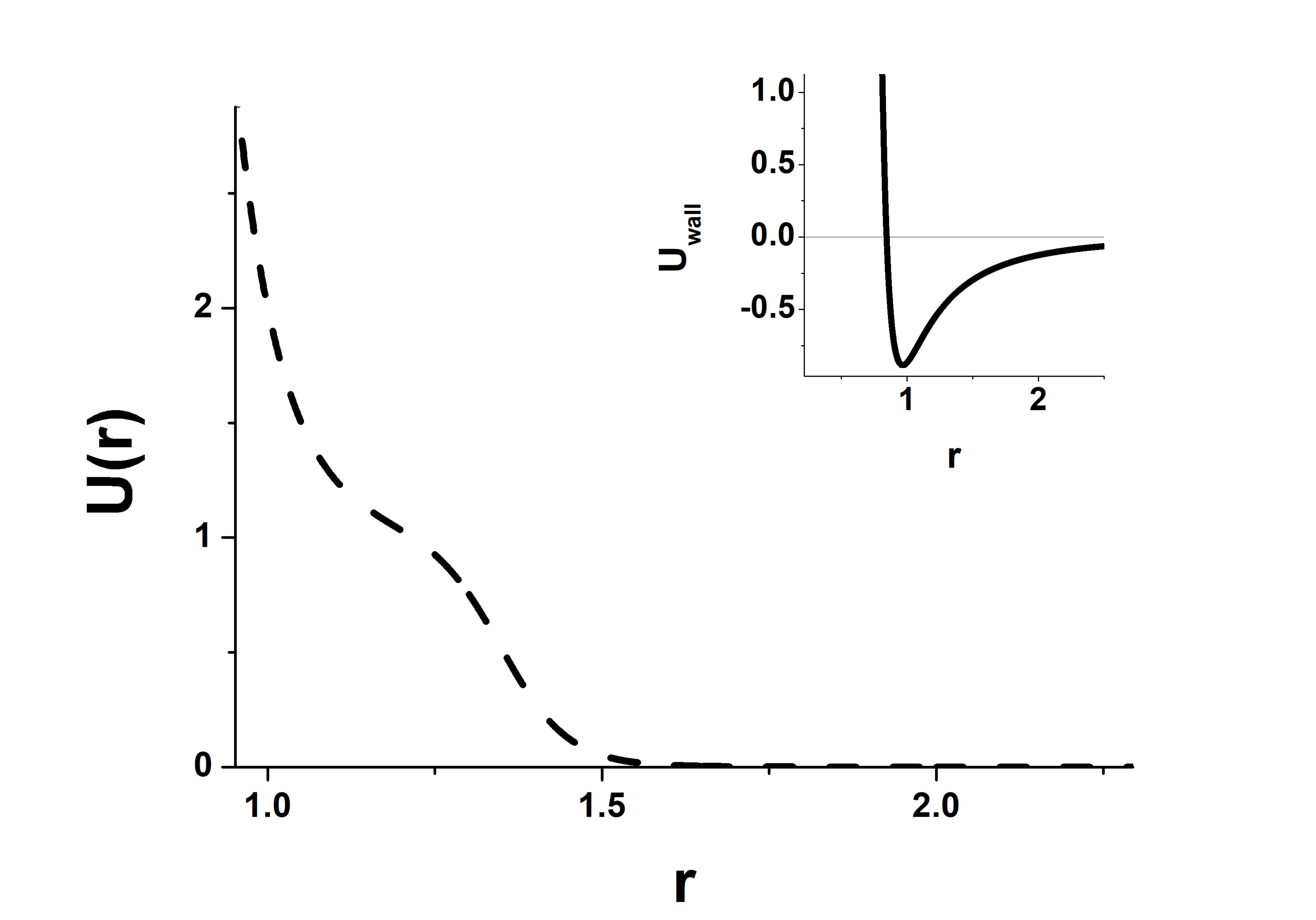}

\caption{The potential (\ref{pot}). The inset shows the interaction between the particles and the walls
from Eq. (\ref{wall}).}
\label{bulk}
\end{figure}

Here we are interested in a water-like density anomaly of the fluid and
influence of the confinement on this anomaly. We study the system
with potential (\ref{pot}) in a slit pore. The walls of the pore are
structureless planes which interact with the particles by the Lennard-Jones 9-3 potential:

\begin{equation} \label{wall}
U_{wall}=\varepsilon \left( \frac{2}{15} \left( \frac{\sigma}{r}
\right)^9  - \left( \frac{\sigma}{r} \right)^3 \right).
\end{equation}
The parameters of the wall-fluid interaction are $\varepsilon =
1.0$, $\sigma = 1.0$ and $r_c=2.5$.

Firstly, we simulated a bulk system of 4000 particles in a cubic
box with periodic boundary conditions. The system was simulated by
means of a molecular dynamics method in canonical ensemble (constant number of particles N,
volume V and temperature T). The time step was set to
$dt=0.0001$. The system was equilibrated for $1 \cdot 10^6$
steps and then more $5 \cdot 10^6$ steps were performed for
calculation of the properties of the system. The density of the system was set to $\rho=N/V=0.5$.
The dependence of pressure on temperature along this isochor
demonstrates a minimum. This minimum is the temperature of the
density anomaly. The temperatures studied in the present work
extend from $T_{min}=0.16$ up to $T_{max}=0.32$ with a step
$\Delta T$=0.02.

After that the system was placed in a slit pore: structureless
walls were placed at the bottom and top of the system. An equilibrated high temperature structure
was used as an initial configuration. The distance between the walls was
selected to be from $H_{min}=10$ up to $H_{max}=40$. The size of
the system in X and Y direction was calculated such that
$L_x=L_y=(N/ \rho)^{1/2}$, where $\rho=0.5$ and N is the number of
particles. N was set to 8000 for the largest pore ($H=40$), 6000
for $H=30$ and $N=4000$ for all smaller pores. All the rest of the
simulation setup is the same as in the bulk case.

In case of bulk system there is a single pressure
$P=\frac{1}{3}(P_{xx}+P_{yy}+P_{zz})$. In case of the system in a
slit pore two pressures are calculated:
$P_{||}=\frac{1}{2}(P_{xx}+P_{yy})$ and $P_{zz}$.

The structure of the system was investigated by calculation of the
density profiles along the Z axis: $\rho (z)=
\frac{N(z+dz)-N(z)}{dz}$, where $N(z)$ is the number of particles
in the flat layers from bottom up to z. In the case of crystal one observes
a set of sharp peaks separated by the gaps with nearly zero density. In the
case of liquid the peaks are much more spread and the density is not zero at any point.

%Typically liquids form one or two layers
%next to the walls while the middle of the system remains liquid-like. In the case of
%crystal the whole system is divided into well defined layers.

%This parameter allows us
%to see if the system forms several layers with liquid-like
%structure or if it crystallizes. Moreover, the structure of the
%system in each layer was studied by calculation of two-dimensional
%radial distribution function $g(r)$.

All simulations were performed using LAMMPS simulation package
\cite{lammps}.

%\section{Results and discussion}

Fig. \ref{bulk} shows the equation of state of the system at an isochor $\rho=0.5$. In accordance to our previous studies the
system demonstrates the density anomaly, i.e. a minimum of the dependence of the pressure
on temperature. In order
to find the location of the minimum we fitted the obtained equation
of state by a polynomial. The temperature of minimum is
$T_{DA}=0.2$.

\begin{figure}
\includegraphics[width=6cm, height=6cm]{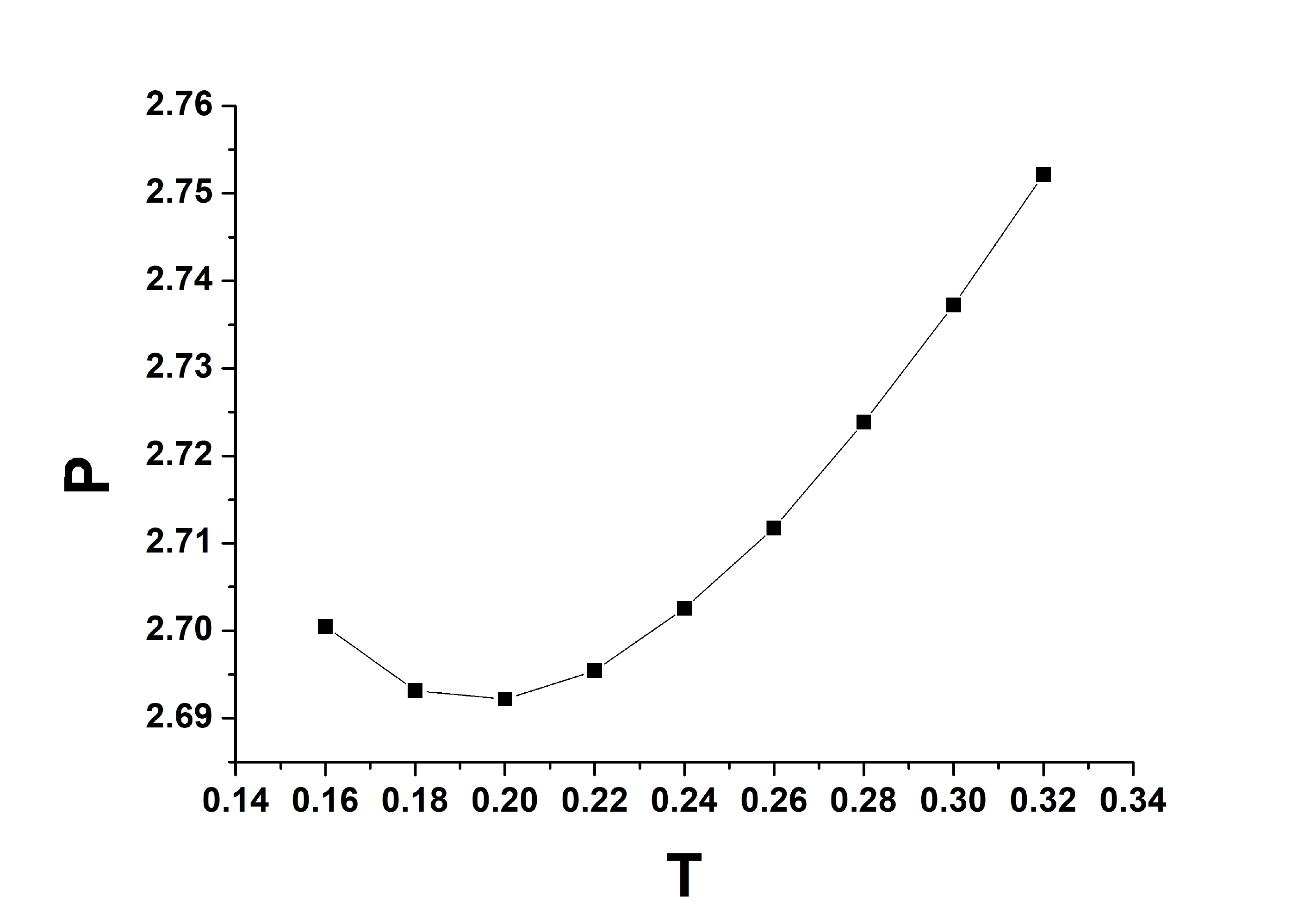}

\caption{The equation of state of the bulk system at $\rho=0.5$.}
\label{bulk}
\end{figure}

\begin{figure}
\includegraphics[width=4cm, height=4cm]{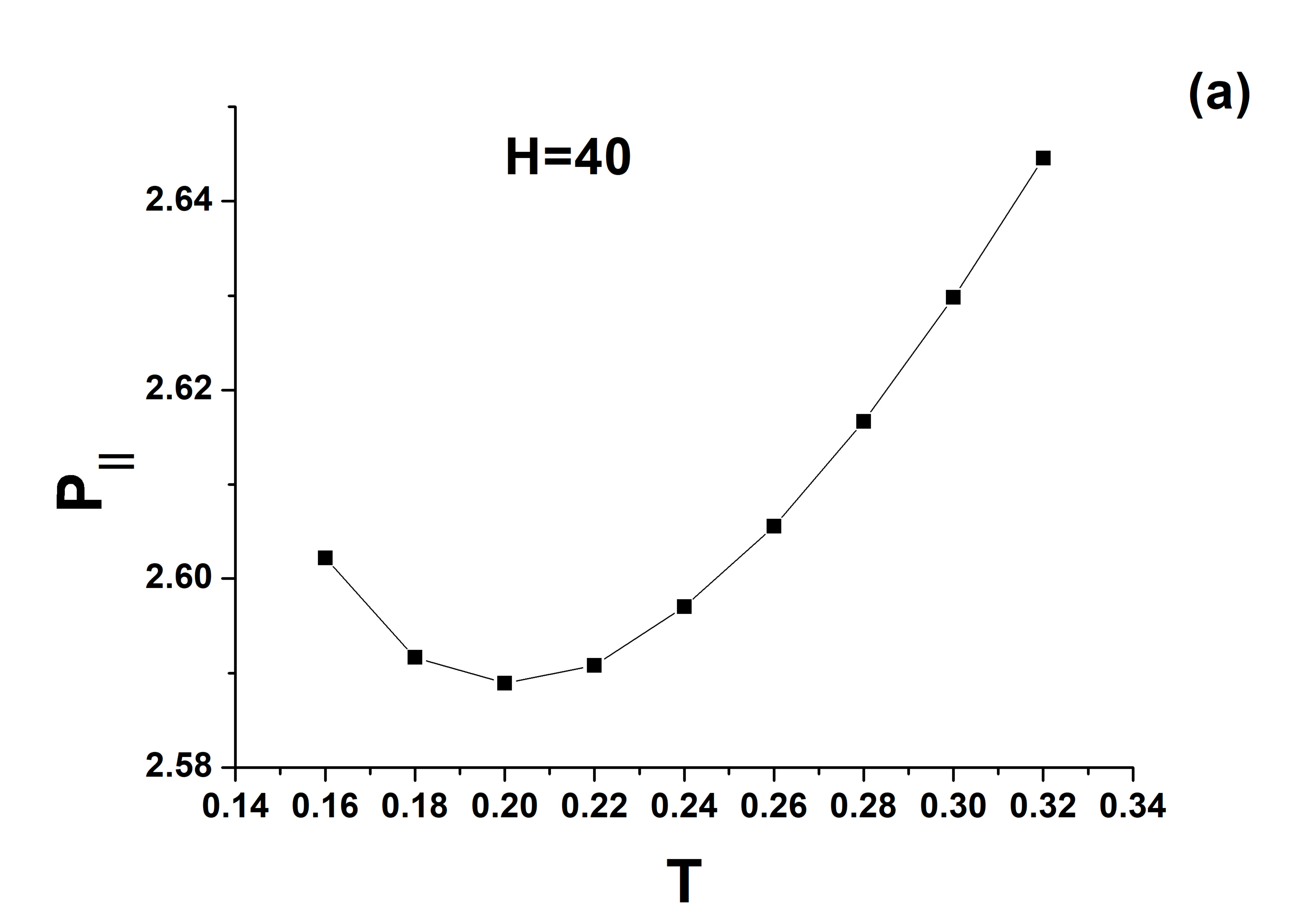}
\includegraphics[width=4cm, height=4cm]{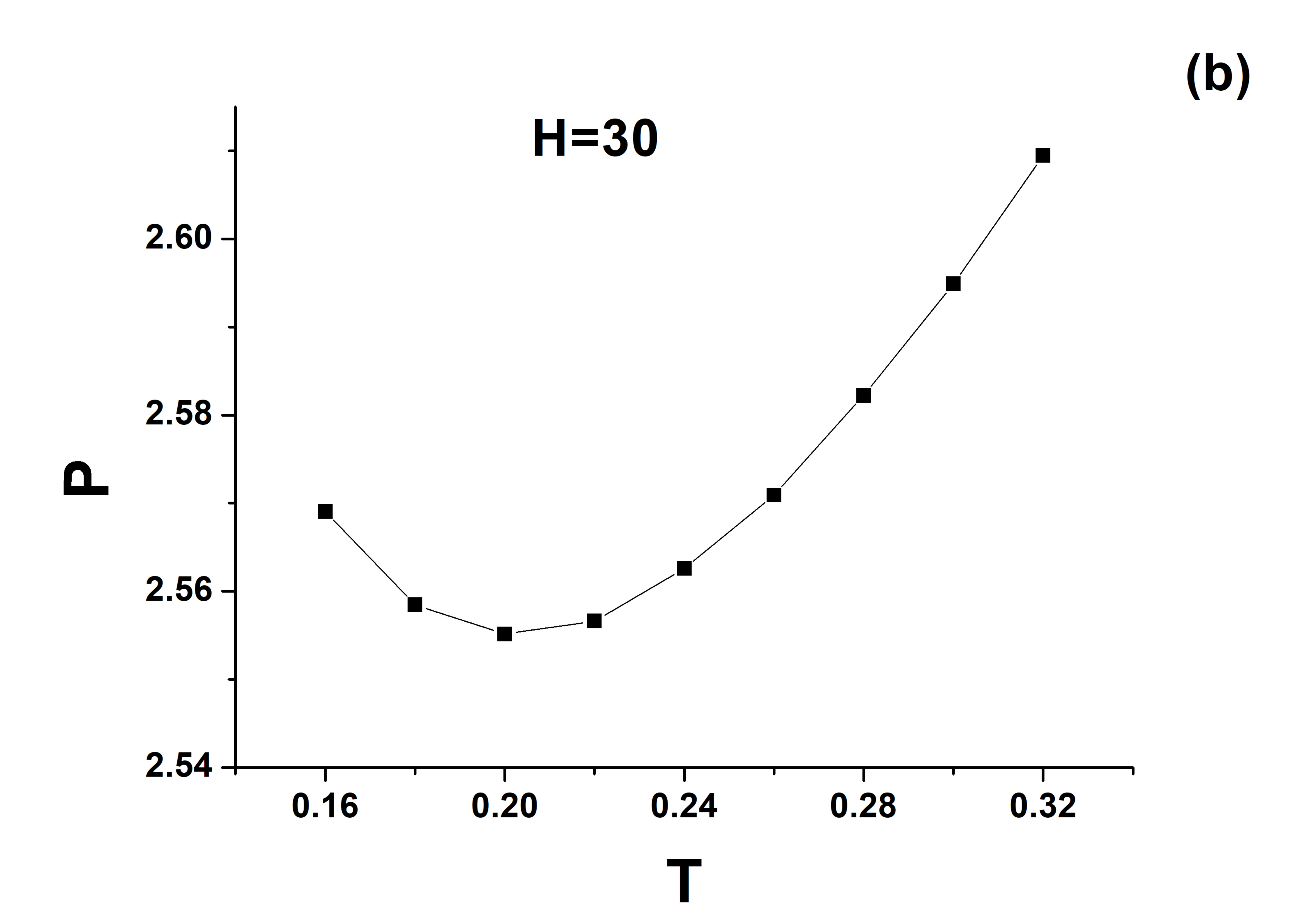}

\includegraphics[width=4cm, height=4cm]{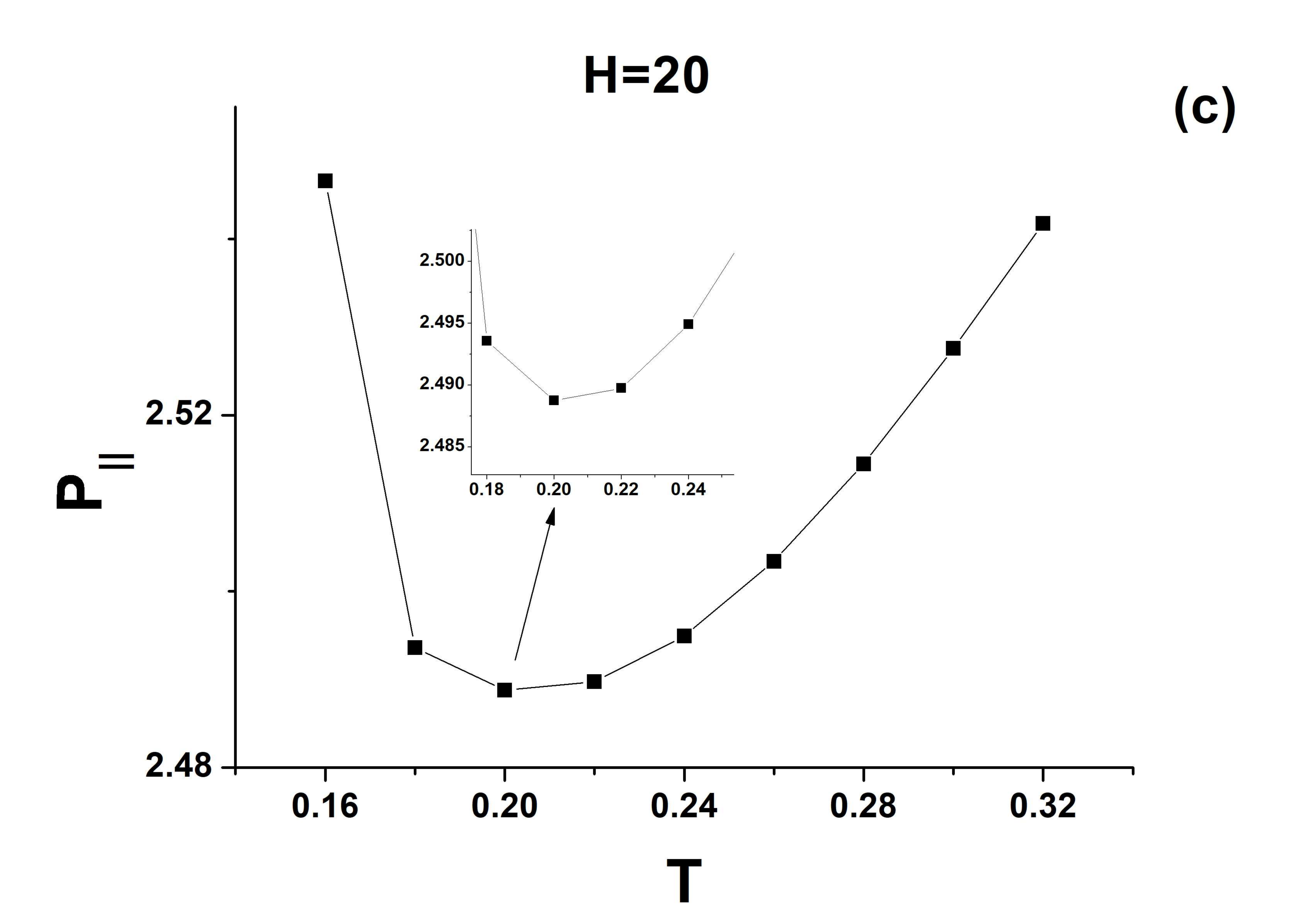}
\includegraphics[width=4cm, height=4cm]{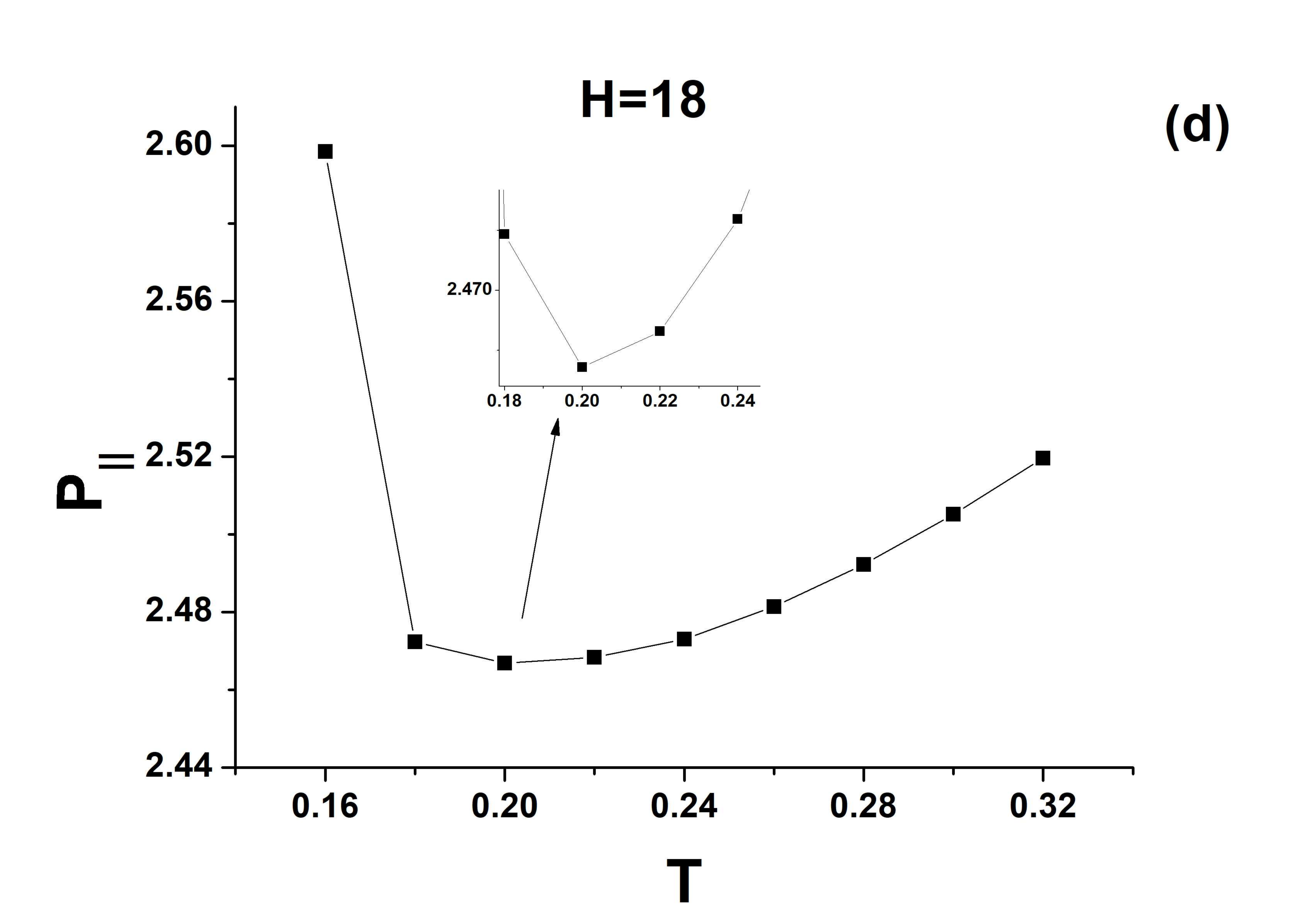}

\includegraphics[width=4cm, height=4cm]{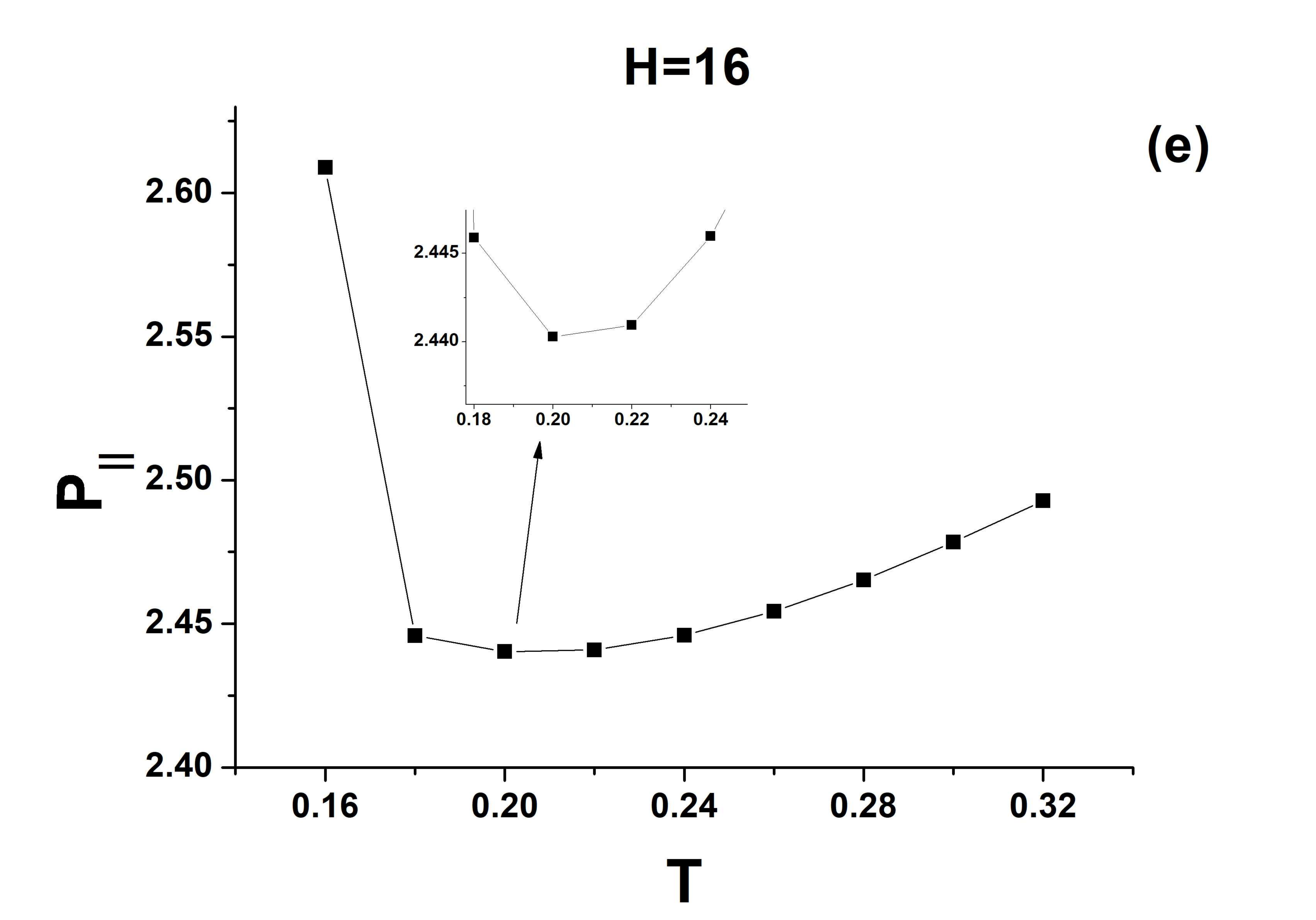}
\includegraphics[width=4cm, height=4cm]{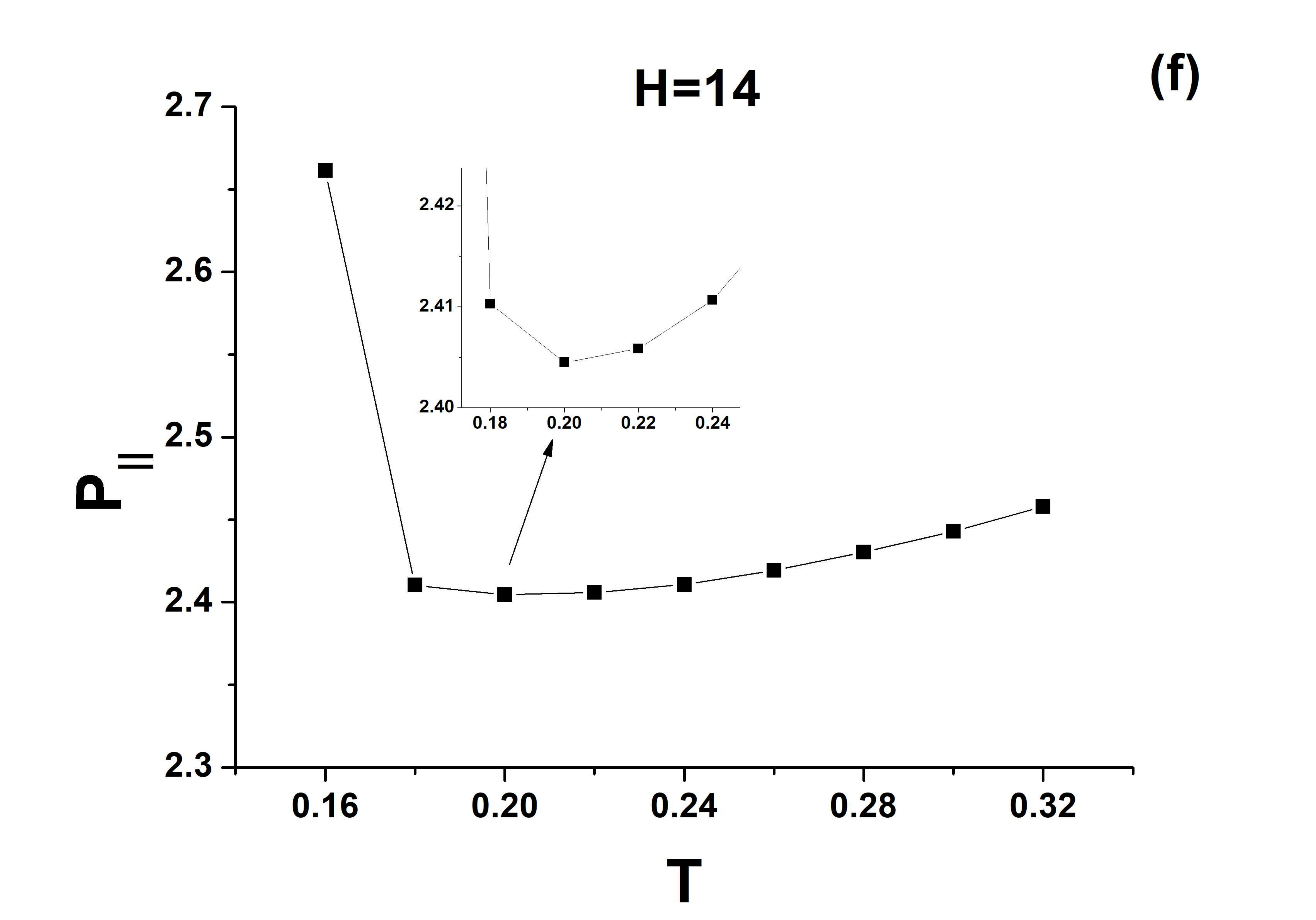}

\includegraphics[width=4cm, height=4cm]{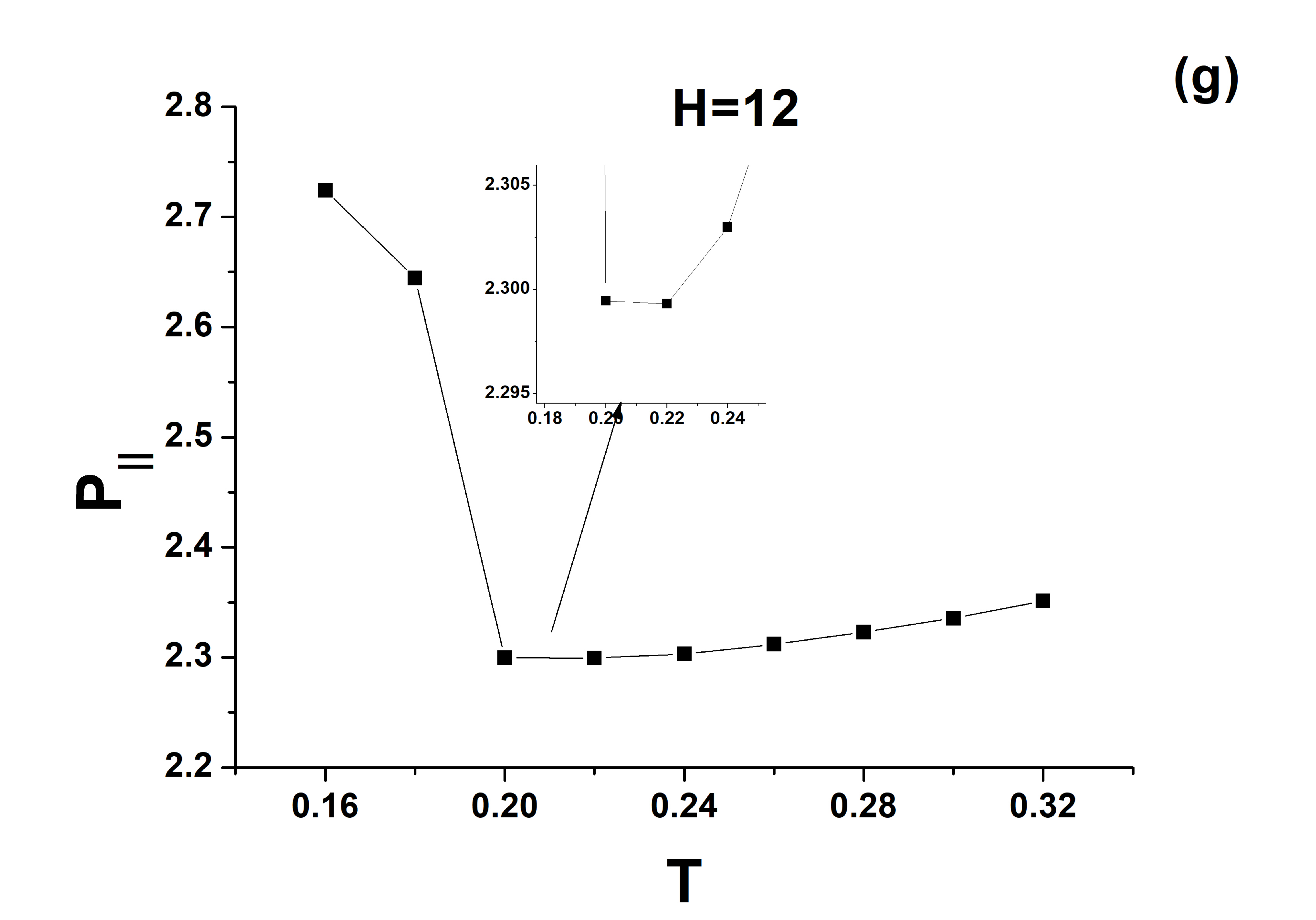}
\includegraphics[width=4cm, height=4cm]{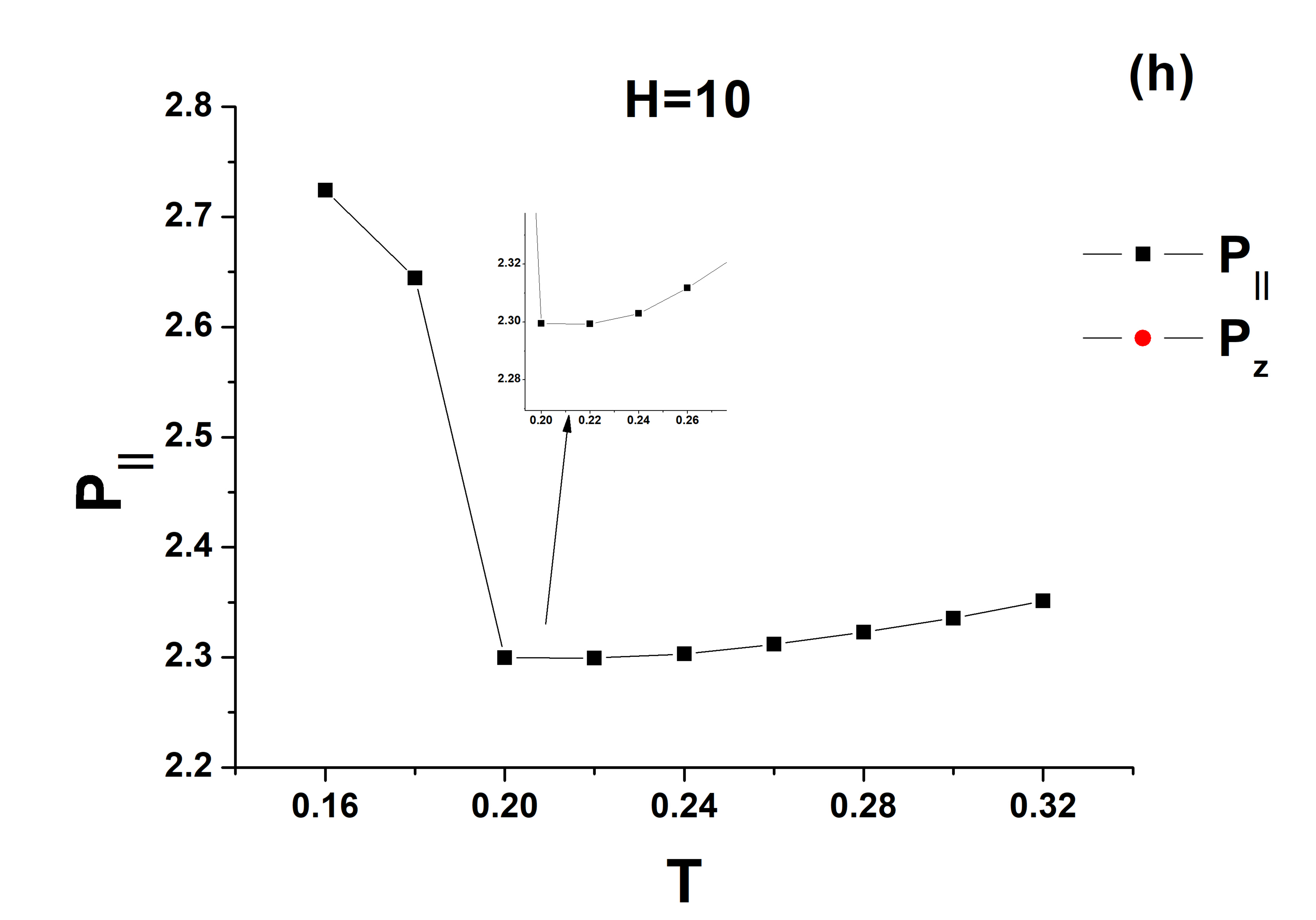}

\caption{The equation of state of the confined systems at different width of the
slit pore. The insets in some panels enlarge the region of temperatures where
density anomaly can be expected.}
\label{conf-eos}
\end{figure}

Fig. \ref{conf-eos} shows the equations of state of the system in slit pores of different width.
We defined the density anomaly via the minimum of in-plane pressure dependence on the temperature $P_{||}(T)$.
One can see that for $H=30$ and $40$ the temperature dependence of pressure is qualitatively similar to the bulk case.
In the pores of smaller width there is a jump of pressure at lowest temperatures which signalizes the crystallization of the system.
However, for $H \geq 14$ we still observe a minimum on the equation of state, i.e. the density anomaly. The temperature
of minimum pressure does not demonstrate any dependence on the pore width: for all cases it is $T_{DA}=0.2$. At the same
time the melting temperature increases with decrease of $H$. While at largest pores ($H=30$ and $40$) we do not observe
crystallization at all studied temperatures, for the pores from $H=14$ to $H=20$ the melting temperature is about $0.16$ and at lowest pores ($H=10$ and $12$)
it reaches about $T_m=0.18$-$0.2$, i.e. it comes very close to the temperature of density anomaly.
One can conclude that the density anomaly is preserved in the confined system, but at strong confinement the melting temperature
becomes higher and approaches the $T_{DA}$. It leads to the depression of the density anomaly. However, a tiny
density anomaly preserves in all studied systems. Moreover, even in purely two-dimensional systems the density anomaly takes place \cite{soft-matt, jcp-we,conf-1layer}.

\begin{figure}
\includegraphics[width=4cm, height=4cm]{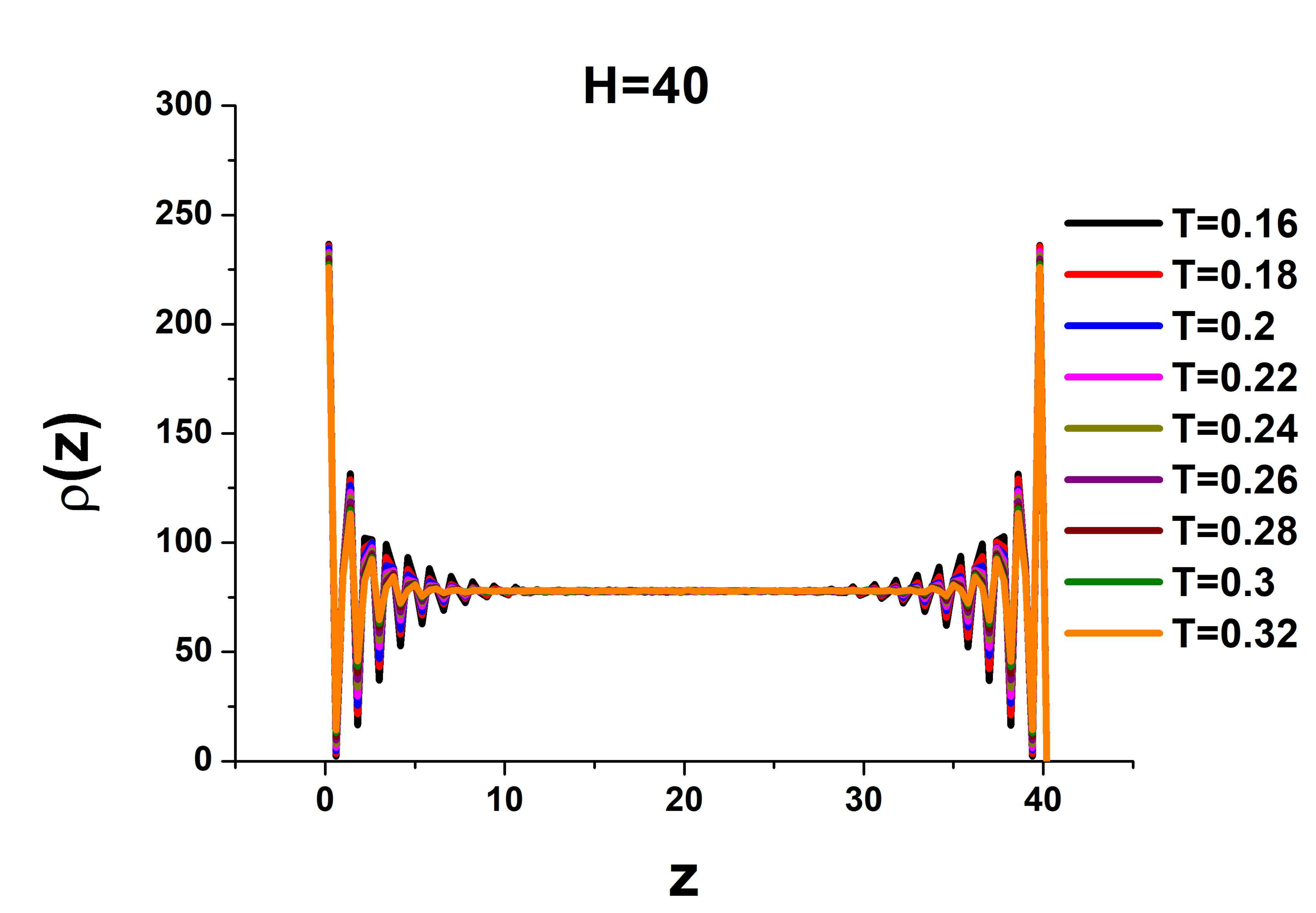}
\includegraphics[width=4cm, height=4cm]{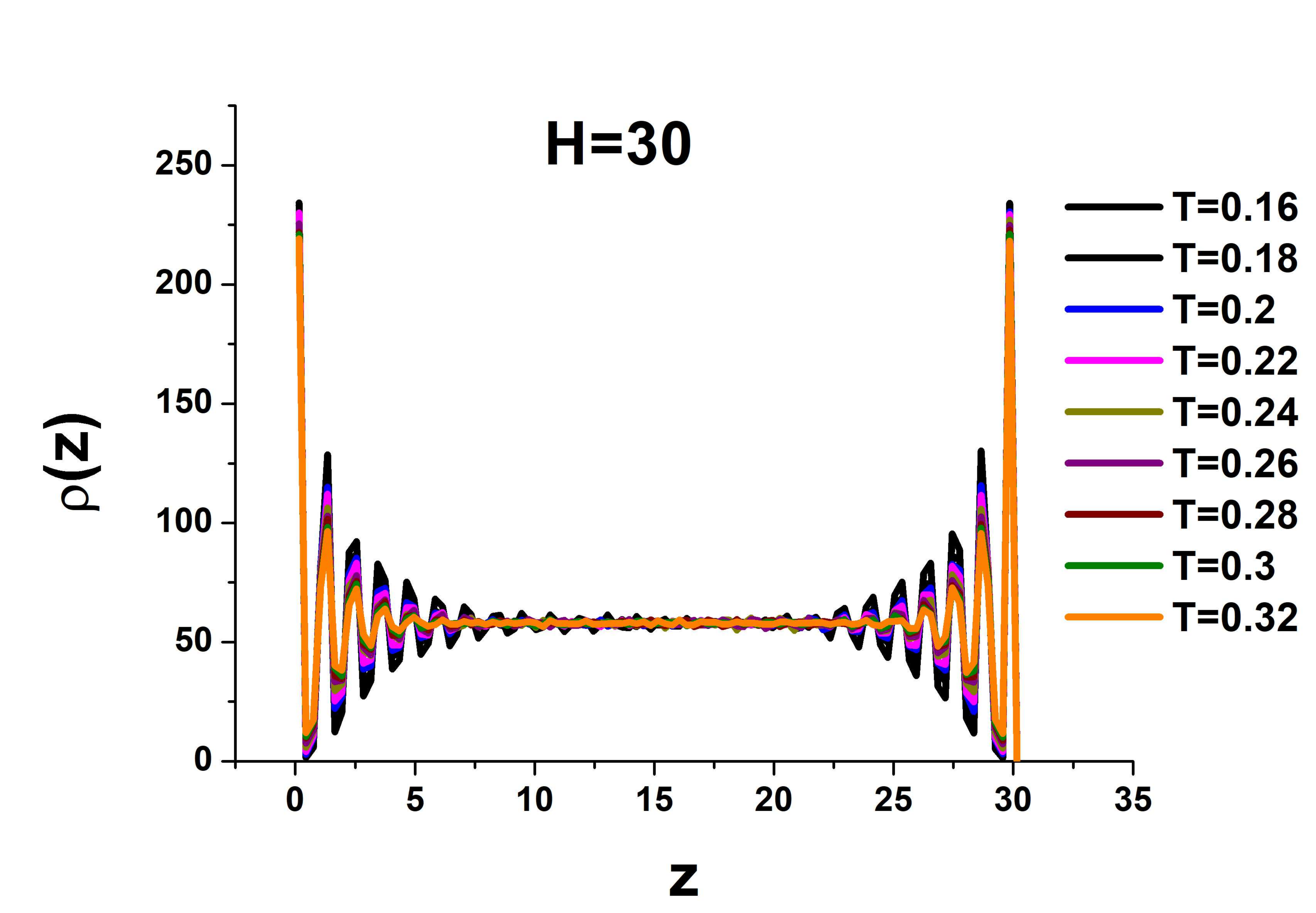}

\includegraphics[width=4cm, height=4cm]{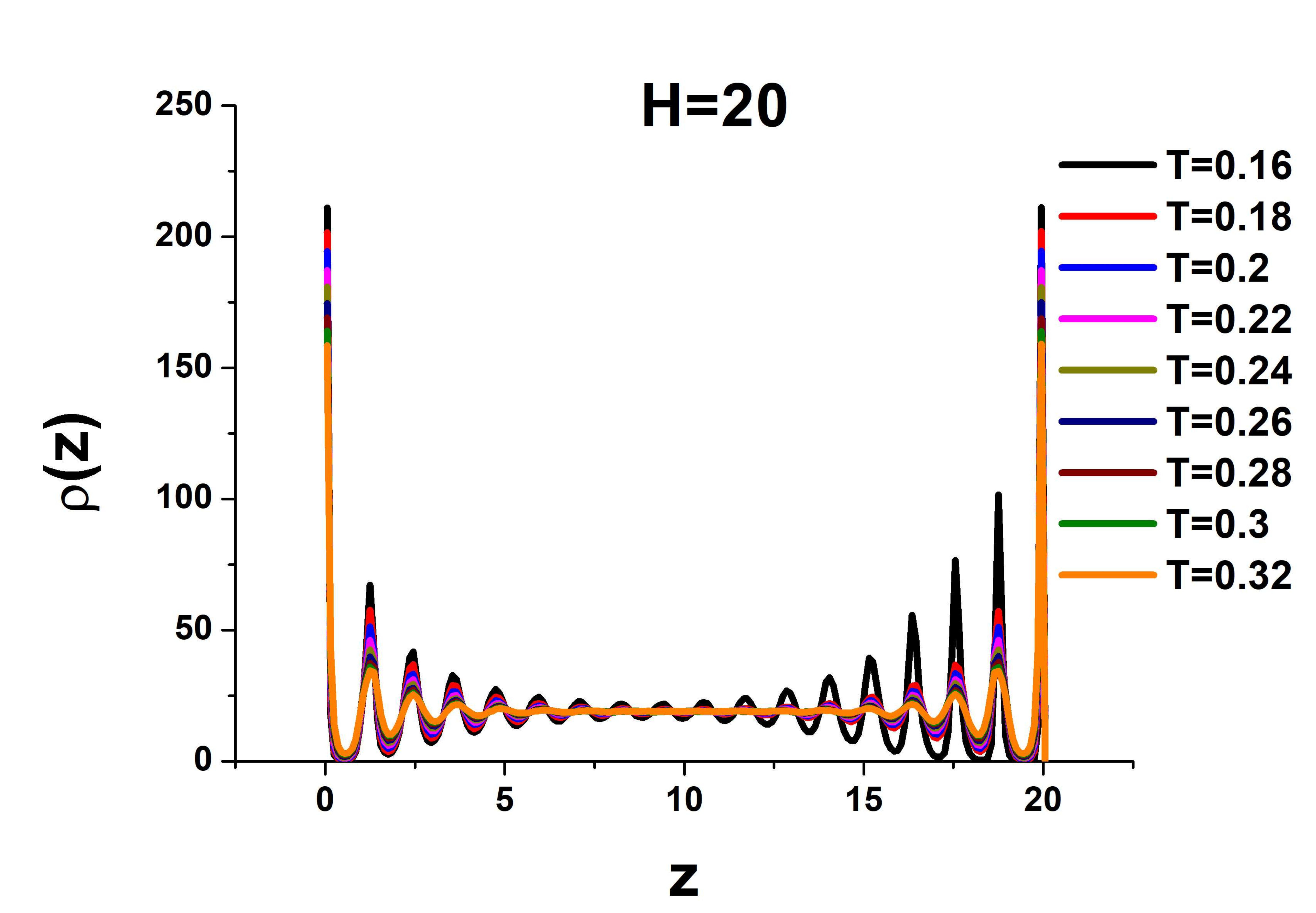}
\includegraphics[width=4cm, height=4cm]{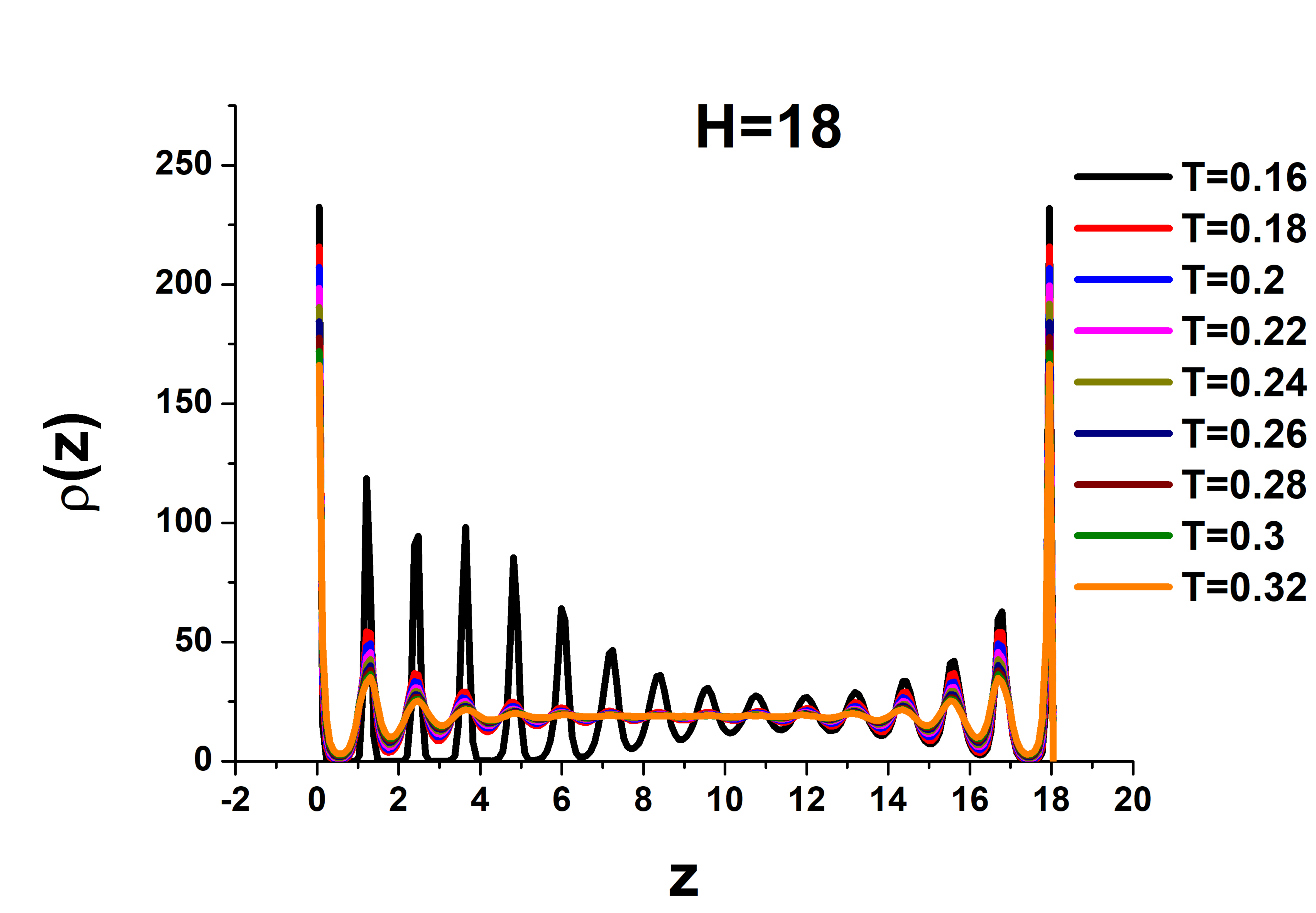}

\includegraphics[width=4cm, height=4cm]{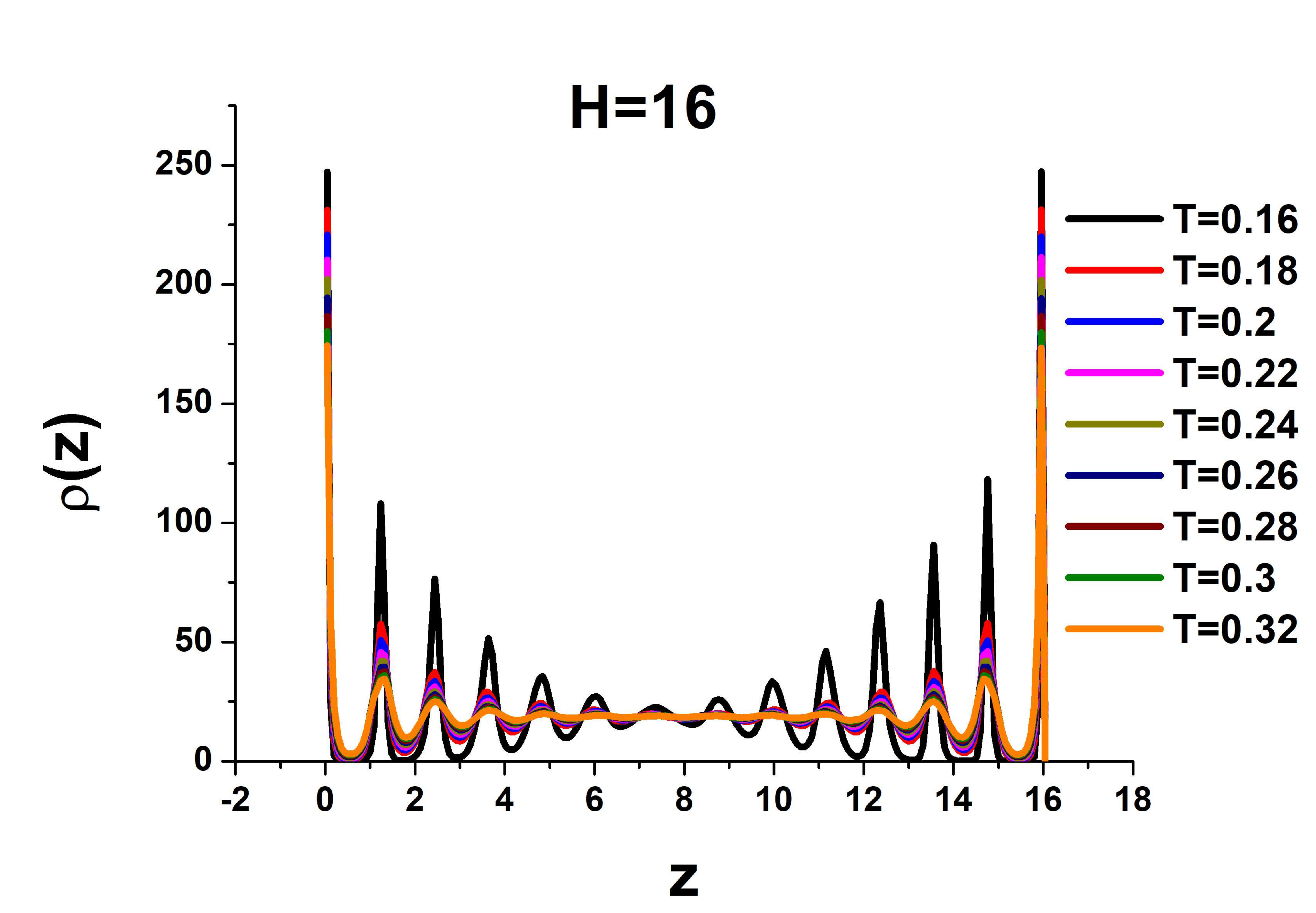}
\includegraphics[width=4cm, height=4cm]{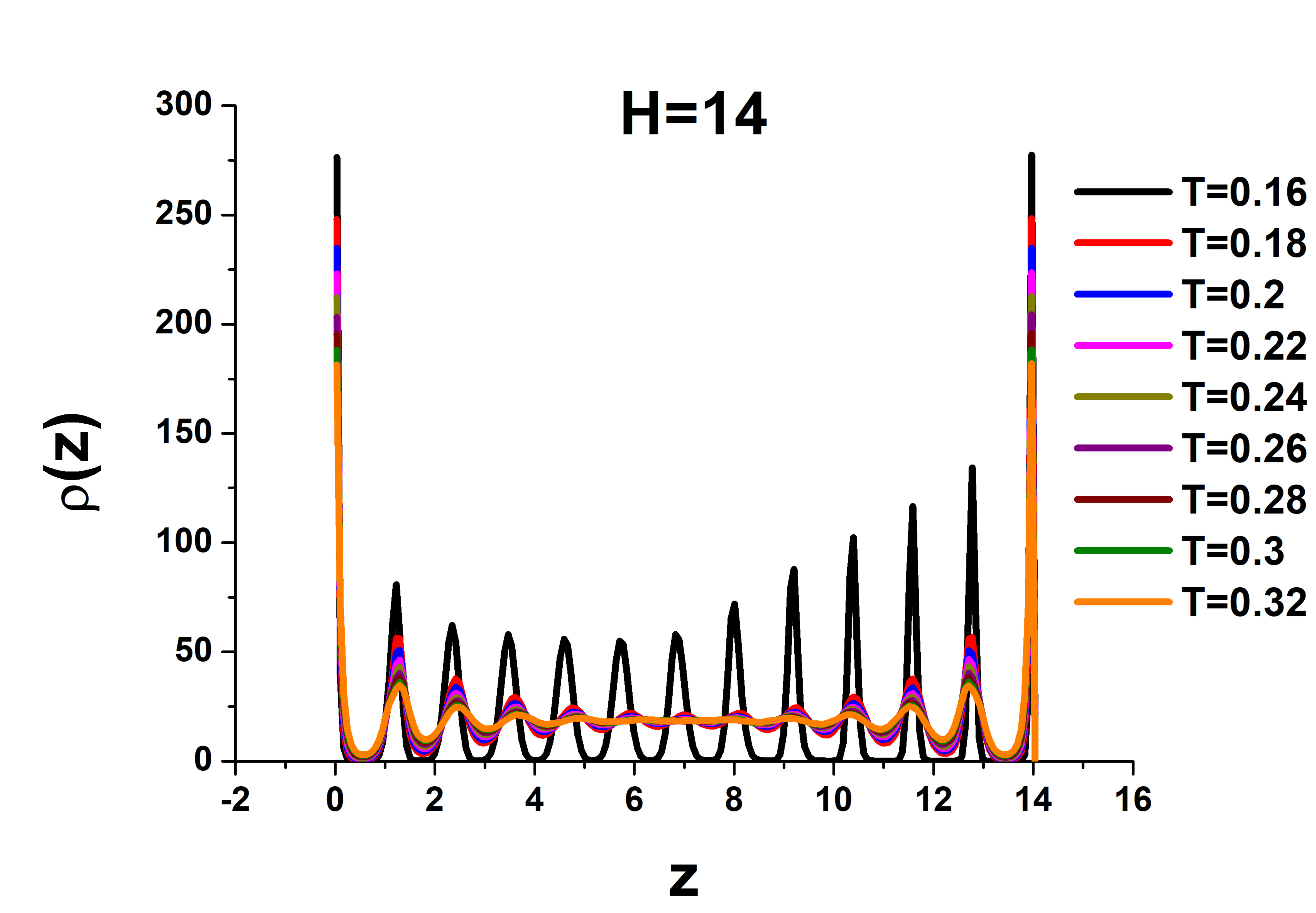}

\includegraphics[width=4cm, height=4cm]{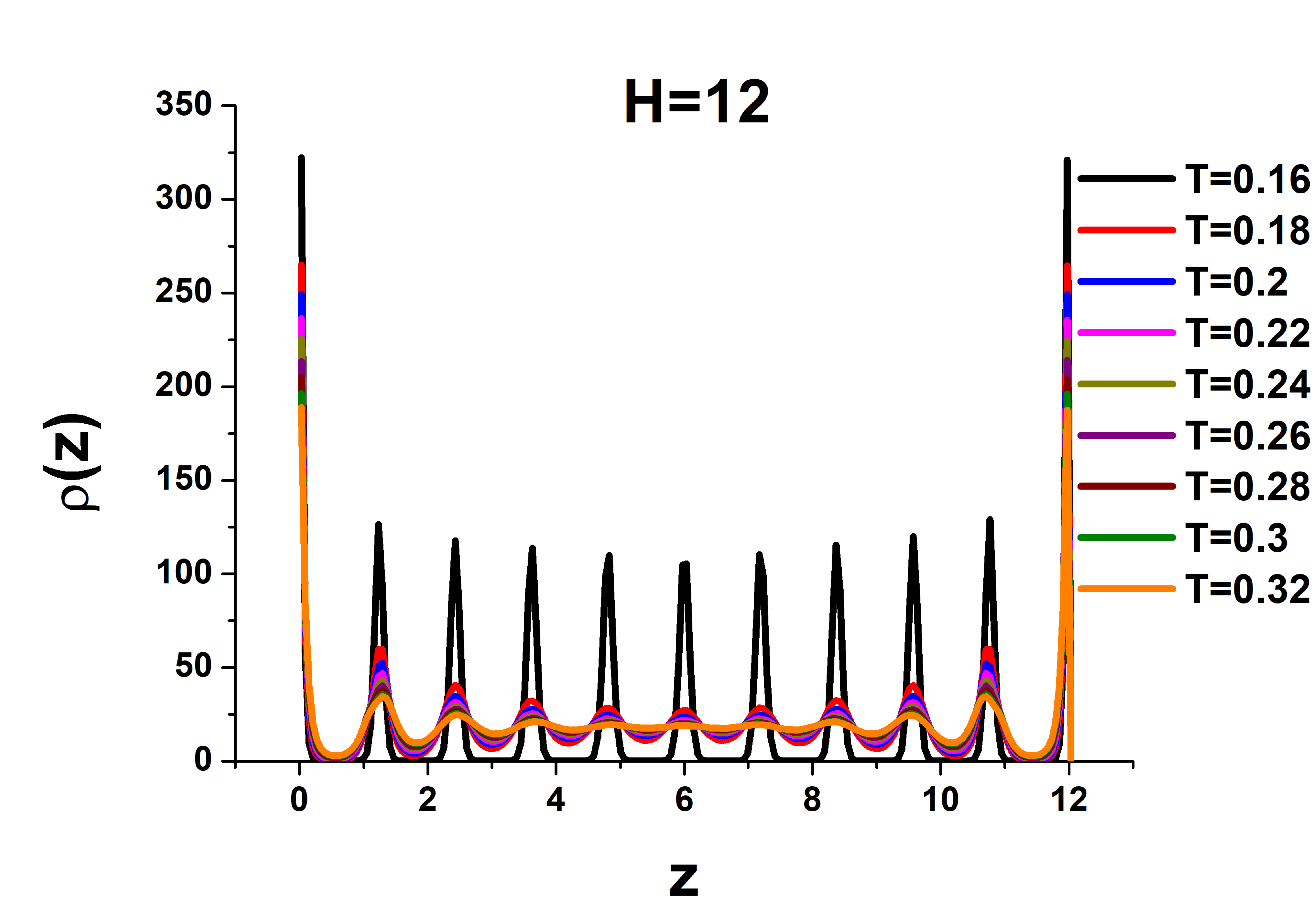}
\includegraphics[width=4cm, height=4cm]{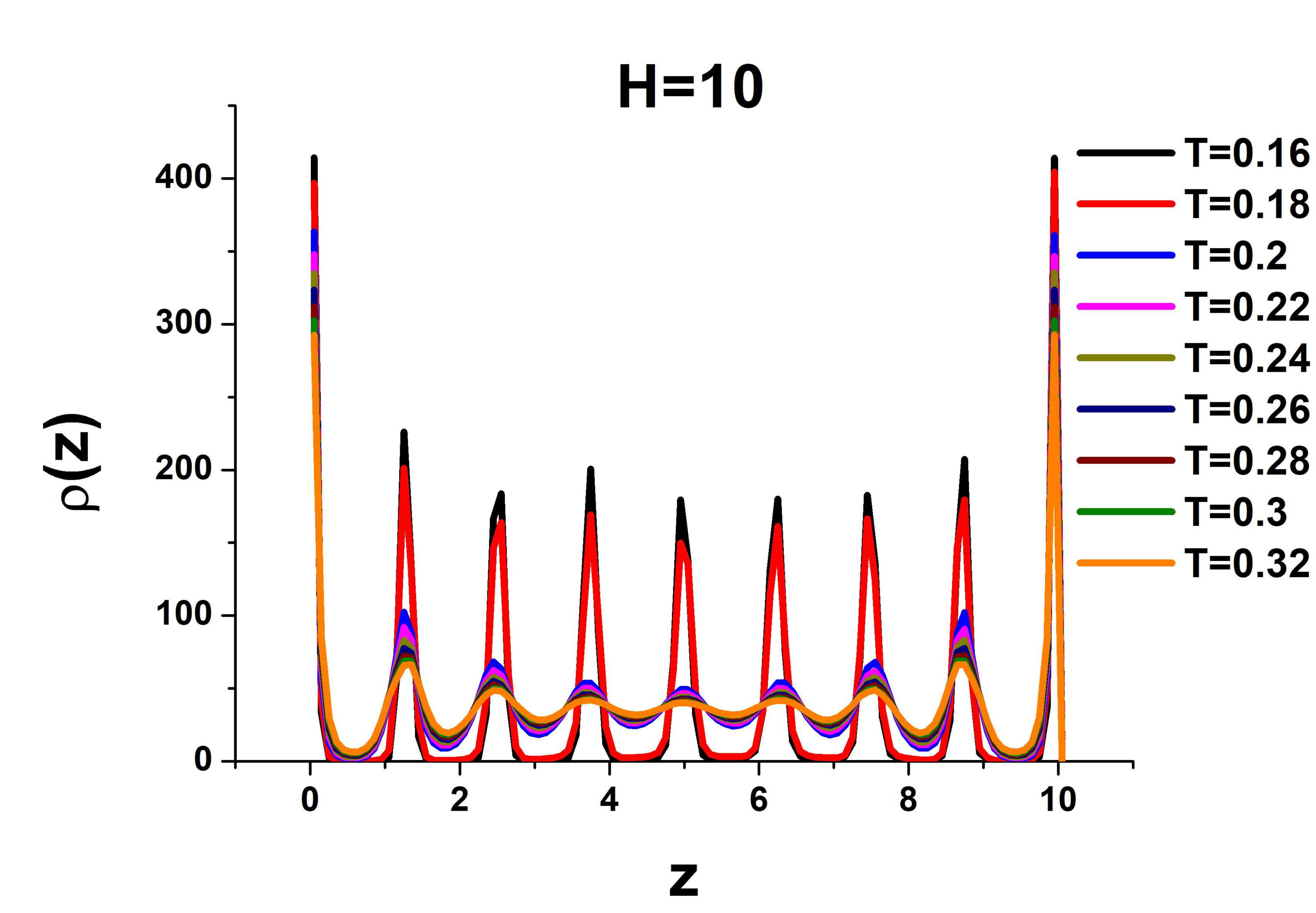}

\caption{The distribution of density from bottom of the slit pore to the top for
all studied widthes of the pore. The widths of the pore is given in each plot.}
\label{conf-rho}
\end{figure}

\begin{figure}
\includegraphics[width=6cm, height=4cm]{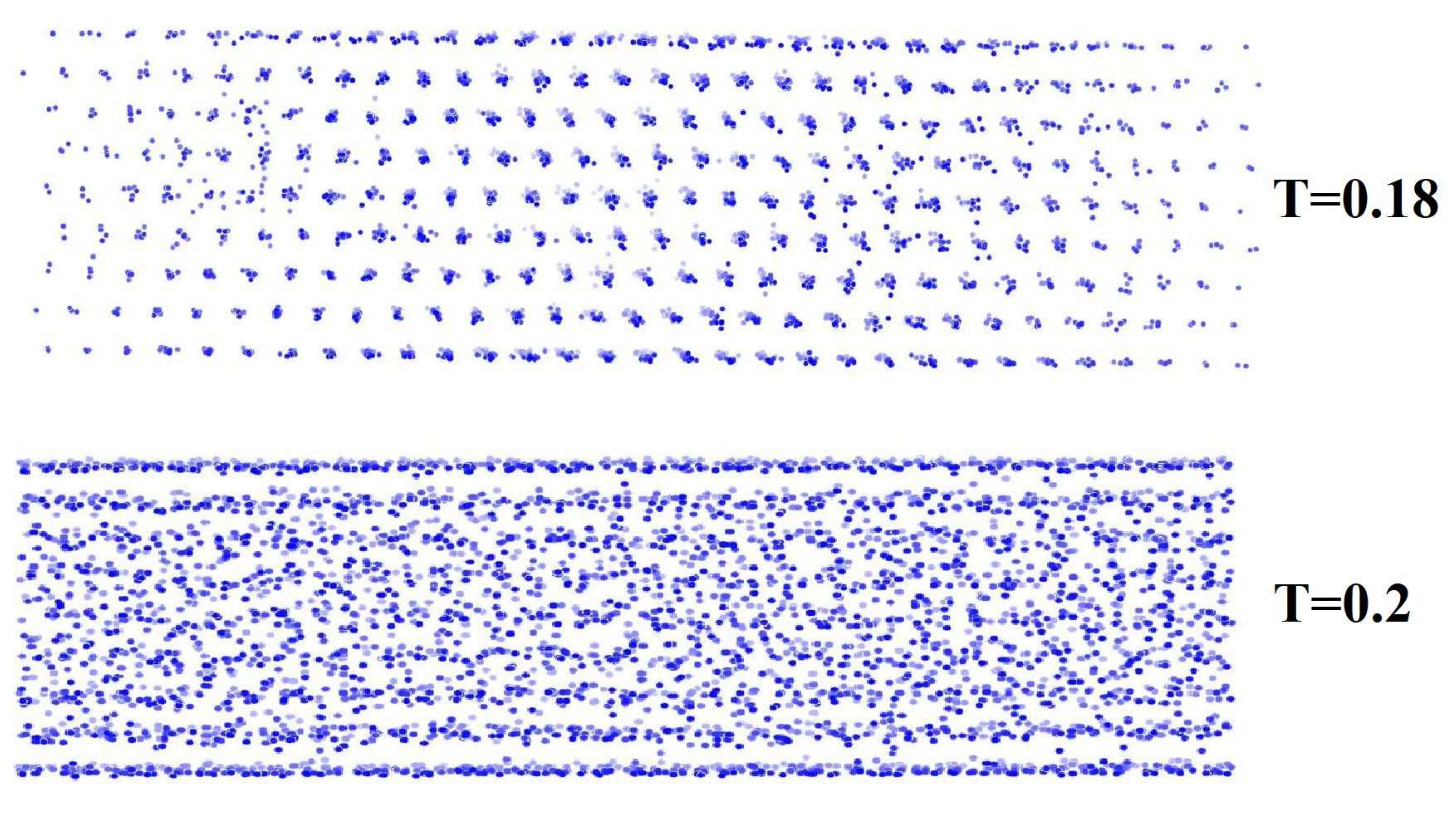}

\caption{Snapshots of the system with $H=10$ at $T=0.18$ (top) and $T=0.2$ (bottom). }
\label{conf-conf}
\end{figure}

The conclusion above is supported by the calculation of density distribution along the z-axis shown in Fig. \ref{conf-rho}.
One can see that at high temperature the system does not demonstrate layering across the whole pore
even with decrease of temperature, which means that although there are some density modulation next to the walls
the system remains liquid. In the case of $H=20$ the layering through the whole pore appears at $T=0.16$, i.e. the system
crystallizes. The system demonstrates two crystalline profiles at $T=0.16$ and $0.18$ with $H=10$,
but at T=0.2 the system transforms to liquid as shown in Fig. \ref{conf-conf}.

\begin{figure}
\includegraphics[width=6cm, height=4cm]{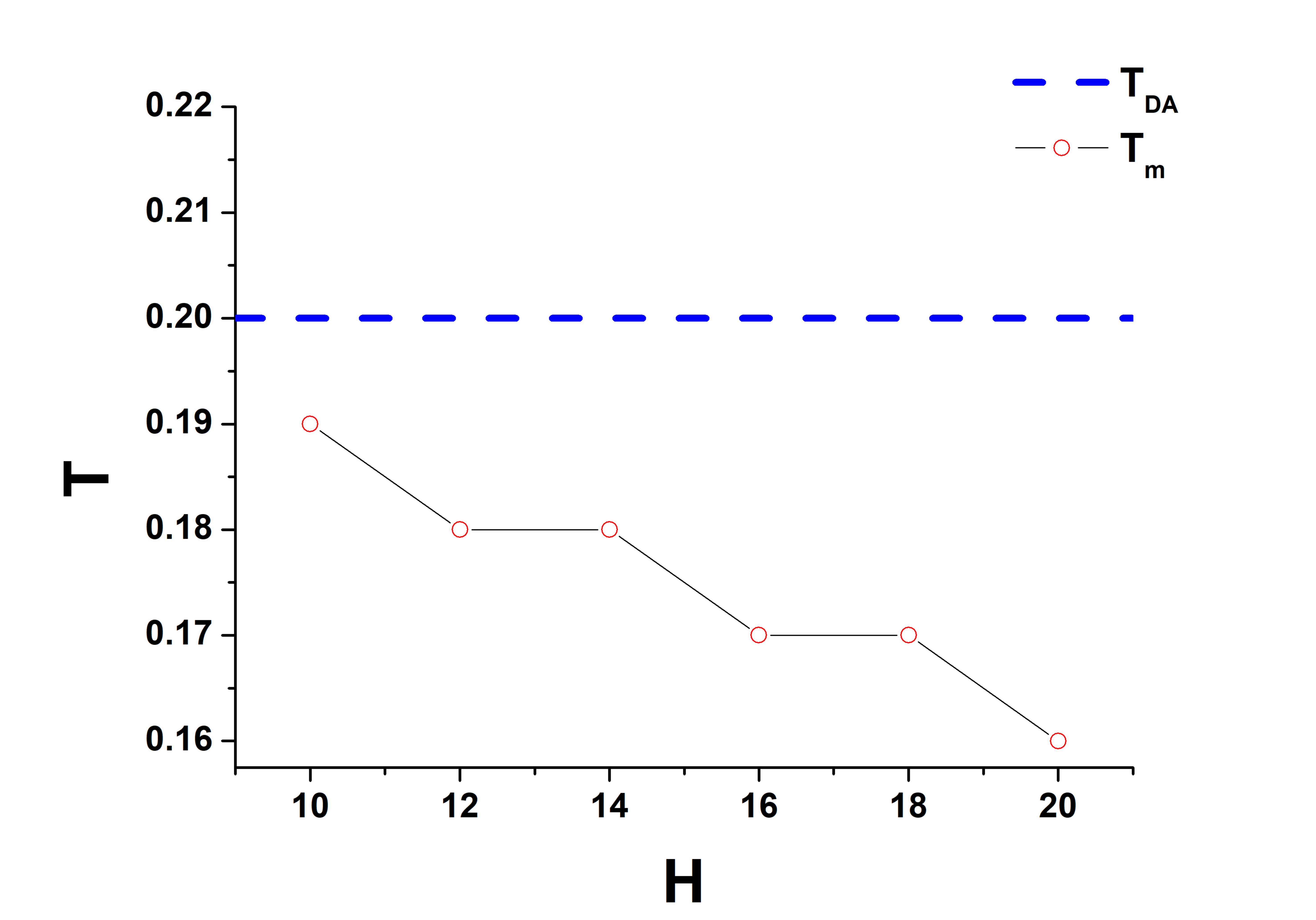}

\caption{Melting temperature of the systems with different width of the pore. Temperature
of the density anomaly is shown for comparison.}
\label{tmtda}
\end{figure}

We estimated the melting temperature from the density profiles of the system. The results are shown in Fig. \ref{tmtda}.
One can see that the melting temperature decreases with an increase in the pore width. However, even with the smallest pores it does not
reach the value of $T_{DA}$. At the same time from the density profiles in Fig. \ref{conf-rho} one can see that
when the temperature of the confined liquid approaches the melting one the layering of the system becomes more pronounced.
The density anomaly is related to some peculiarities of the microscopic structure of a liquid. In the case of
strongly confined fluid which we do observe in narrow pores at low temperatures the density is strongly influenced by the
walls. The local density distribution becomes determined mostly by the particle-wall interaction, which leads to the depression
of the density anomaly.

%which leads to the disappearance of the density anomaly.

\bigskip

In conclusion, we performed a study of  the density anomaly of a liquid confined in a slit pore. We found that the
temperature of minimum of pressure dependence on temperature along an isochor did not depend
on the pore width. However, in narrow pores the liquid crystallizes and the crystallization
temperature increases with decrease of the pore width. As a result the local density is mostly controlled by
the particle-wall interaction, which is does not produce the density anomaly. This result is important for
understanding of the behavior of anomalous liquids in confined medium, in particular, water.

%At the temperatures slightly above the melting
%one the liquid demonstrates strong layering which leads to disappearance of the density anomaly.
%This result is important for understanding of the behavior of anomalous liquids
%in confined medium, in particular, water.

This work was carried out using computing resources of the federal
collective usage center "Complex for simulation and data
processing for mega-science facilities" at NRC "Kurchatov
Institute", http://ckp.nrcki.ru, and supercomputers at Joint
Supercomputer Center of the Russian Academy of Sciences (JSCC
RAS). The work was supported by the Russian Science
Foundation (Grants No 19-12-00092).

\end{document}